\documentclass[12pt]{article}

\def\d{\delta}
\def\L{\Lambda}
\def\l{\lambda}

\def\g{\gamma}
\def\e{\epsilon}

\def\f{\phi}
\def\t{\tau}
\def\o{\omega}
\def\O{\Omega}
\def\g{\gamma}
\def\G{\Gamma}

\def\th{\theta}
\def\c{\chi}
\def\k{\kappa}

\def\pr{\partial}

\def\ha{\frac{1}{2}}
\newcommand{\be}{\begin{equation}}
\newcommand{\ee}{\end{equation}}
\newcommand{\bea}{\begin{eqnarray}}
\newcommand{\eea}{\end{eqnarray}}

\newcommand{\R}{{\sf R\hspace*{-0.9ex}\rule{0.15ex}%
{1.5ex}\hspace*{0.9ex}}}
\def\bkR{{\rm I\kern-.17em R}}
\def\bkC{{\rm \kern.24em \vrule width.05em height1.4ex depth-.05ex \kern-.26em C}}
\def\bkN{{\rm \kern.50em \vrule width.05em height1.4ex depth-.05ex \kern-.26em N}}

\begin{document}
\bigskip
\bigskip
\bigskip
\begin{center}
\Large{\bf Coherent States Expectation Values as Semiclassical Trajectories}
\end{center}
%\markboth{Authors' Names}
%{Instructions for Typing Manuscripts ( )}
\bigskip
\bigskip
\bigskip
\begin{center}
N. C. DIAS\footnote{ncdias@mail.telepac.pt}, A. MIKOVI\'C\footnote{amikovic@ulusofona.pt} and J. N. PRATA\footnote{joao.prata@ulusofona.pt}\\ 
Departamento de Matem\'atica \\
Universidade Lus\'ofona de Humanidades e Tecnologias\\Av. do Campo Grande, 376, 1749-024 Lisboa, Portugal
\end{center}

\bigskip
\bigskip
\begin{center}
{\bf Abstract}
\end{center}
We study the time evolution of the expectation value of the anharmonic oscillator coordinate in a coherent state as a toy model for understanding the semiclassical solutions in quantum field theory. By using the deformation quantization techniques, we show that the coherent state expectation value can be expanded in powers of $\hbar$ such that the zeroth-order term is a classical solution while the first-order correction is given as a phase-space Laplacian acting on the classical solution. This is then compared to the effective action solution for the one-dimensional $\f^4$ perturbative quantum field theory. We find an agreement up to the order $\l\hbar$, where $\l$ is the coupling constant, while at the order $\l^2 \hbar$ there is a disagreement. Hence the coherent state expectation values define an alternative semiclassical dynamics to that of the effective action. The coherent state semiclassical trajectories are exactly computable and they can coincide with the effective action trajectories in the case of two-dimensional integrable field theories. 

\newpage
\section{Introduction}

The notion of a semiclassical trajectory as a quantum corrected classical trajectory is a very useful idea in various areas of physics. In the case of field theories, the classical trajectory represents a classical field configuration, and the semiclassical field configurations are usually calculated from the effective action \cite{schw,jlas}, i.e. by solving the corresponding equations of motion . 

The effective action approach was tailor-made
for the problems of scattering of the elementary particles, and strictly speaking, one is not calculating the expectation value of an appropriate operator, but a matrix element between the ``in" and the ``out" vacuum. This means that the obtained values can be complex and this is a problem in the context of quantum gravity and quantum cosmology applications because the field operator is a metric, and the effective metric has to be real. This problem can be resolved by using the effective action formalism where the field variable is a true expectation value \cite{jord,cahu}. However, in order to obtain a semiclassical trajectory, one needs to solve the corresponding effective equation of motion, which is often a difficult task. 

This type of problems were encountered in the context of two-dimensional dilaton gravity models of quantum black holes, where the effective metric gives the information about the back-reaction of the black hole evaporation \cite{2ddg}. It was demonstrated in \cite{m} that a relevant one-loop ($O(\hbar)$) solution could be obtained as an expectation value of the metric operator in the coherent state corresponding to the initial matter distribution. Furthermore, a two-loop ($O(\hbar^2 )$) solution was found by using this approach \cite{mr}, which was otherwise impossible to do by solving the corresponding two-loop effective action equations.   

This then suggests an approach to constructing semiclassical trajectories for field theories via expectation values of the appropriate operators in the coherent states. This is reasonable because the coherent states are the closest approximation to the classical point in the phase space. Using the coherent states in the perturbative quantum field theory (QFT) can be justified by the fact that the ``in" and the ``out" states correspond to asymptotically free particles. The free particles in QFT can be considered as a collection of harmonic oscillators, and it is known that the expectation value of the harmonic oscillator coordinate in the coherent state follows the classical trajectory.

In this paper we will study the expectation values of the Heisenberg operators in the coherent states for finite-dimensional systems as a preparation for the field theory case. In \cite{jl3} and \cite{br} a study of the classical and the effective quantum dynamics of field theory toy models (anharmonic oscillators) was performed, but in the effective potential approximation. It is possible to improve this approximation by including more relevant terms in the effective action \cite{jl3}, but there is no simple expression for the complete one-loop contribution, i.e. the $\hbar$ correction. However, it has been known from the deformation quantization applications to quantum optics how to evaluate the coherent states expectation values \cite{Lee}. Hence one can apply these techniques to the case of the anharmonic oscillator (AHO). The result can be then expanded perturbatively in the coupling constant and compared to the corresponding effective action expansion.

In section two we introduce the basic concepts of deformation quantization (DQ) and derive a new formula for the expectation value (EV) of a generic dynamical variable in an arbitrary state. We then use this result in section three to derive the $O(\hbar)$
contribution to the coherent state EV of a generic dynamical variable. We then derive the corresponding quantum equation of motion for the coordinate EV and in section four we specialize to the case of an AHO. In section five we derive the perturbative quantum equations of motion for an AHO coming from the effective action formalism and compare the results of the two approaches. In section six we present our conclusions and in the Appendix we describe how to solve the perturbative equations of motion. 
  
\section{Expectation values} 

Let us consider an $N$-dimensional system with coordinates $q= \left(q_1, \cdots , q_N \right)$ and canonical momenta $p=\left(p_1, \cdots , p_N \right)$. We shall assume a flat phase space $T^* M \simeq \R^{2N}$ with symplectic form $\sigma (z, z') = q \cdot p' - p \cdot q'$, where $z= (p,q)$, $z'= (p',q')$. 

In the context of deformation quantization \cite{Lee}-\cite{Narcowich2} one computes the expectation value of a generic operator $\hat A \left( \hat z ,t \right)$ from the algebra of observables at time $t$ in a state $\psi \in L^2 \left( \R^N, dq \right)$ as
\begin{equation}
A(t) \equiv \langle \psi| \hat A ( \hat z, t) | \psi\rangle = \int dz  \, F_W (z) A_W (z,t)\,,
\end{equation}  
where $F_W(z)$ is the Wigner function \cite{Wigner} associated with $\psi$
\begin{equation}
F_W (p,q) = \frac{1}{\left(\pi \hbar\right)^N} \int dy \, e^{-2ip \cdot y / \hbar} \psi^* (q-y) \psi (q+y)\,.
\end{equation}
$A_W(z,t)$ is the Weyl symbol \cite{Weyl} associated with $\hat A (\hat z,t)$ given by
\begin{equation}
A_W (z,t) = \left( \frac{\hbar}{2 \pi} \right)^N \int d \xi \, Tr \left\{\hat A ( \hat z ,t ) e^{i \xi \cdot \hat z} \right\} e^{- i \xi \cdot z}\,.
\end{equation}  
The Weyl symbol defines a noncommutative twisted product \cite{Groenewold},
\begin{equation}
\begin{array}{c}
\left(\hat A \cdot \hat B \right)_W \equiv \left. A_W \star_W B_W = \exp \left[\frac{i\hbar}{2} \sigma \left( \frac{\partial}{\partial z_1} , \frac{\partial}{\partial z} \right) \right] A_W (z_1) B_W (z) \right|_{z_1 =z} = \\
\\
= A_W (z) B_W (z) + {\cal O} (\hbar)\,,
\end{array}
\end{equation}
where $\frac{\partial}{\partial z} = \left( \frac{\partial}{\partial p} , \frac{\partial}{\partial q} \right)$. Likewise, one may define a bracket- the Moyal bracket - according to \cite{Moyal}:
\begin{equation}
\begin{array}{c}
\left[ A_W, B_W \right]_W \equiv \left(\frac{1}{i \hbar} \left[\hat A , \hat B \right] \right)_W  =
 \frac{1}{i \hbar} \left( A_W \star_W B_W - B_W \star_W A_W \right) =\\
\\
= \left. \frac{2}{\hbar}  \sin \left[\frac{\hbar}{2} \sigma \left( \frac{\partial}{\partial z_1} , \frac{\partial}{\partial z} \right) \right] A_W (z_1) B_W (z) \right|_{z_1 =z} 
= \left\{A_W (z) , B_W (z) \right\} +  O(\hbar^2)\,,
\end{array}
\end{equation}
These algebraic operations are formal deformations of the usual product and of the Poisson bracket with deformation parameter $\hbar$. 

The dynamics is governed by the Moyal equation
\begin{equation}
\dot A_W (z,t) = \left[H_W (z), A_W (z,t) \right]_W =  \left\{H_W (z), A_W (z,t) \right\} +  O (\hbar^2)\,,
\label{six}\end{equation}
where $H_W$ is the Weyl symbol of the quantum Hamiltonian. 

The following remark is important for the sequel. If the Hamiltonian is of the form 
\begin{equation}
H = \frac{p^2}{2m} + U( q)\,,\label{seven}
\end{equation}
then there are no ordering ambiguities and we conclude that eq.(6) only yields corrections of even order in $\hbar$ to the classical solution $A_{cl} (z,t)$ so that
\begin{equation}
A_W (z,t) = A_{cl} (z,t) + O (\hbar^2)\quad.\label{eight}
\end{equation}  

The Wigner function being a square integrable function admits a Fourier transform:
\begin{equation}
\tilde F_W (a) = \int dz \hspace{0.2 cm} F_W (z) e^{i a \cdot z} = \int dx \hspace{0.2 cm} e^{i v \cdot x} \psi^* \left( x - \frac{\hbar u}{2} \right) \psi \left( x + \frac{\hbar u}{2} \right)  ,
\end{equation}
with the inverse
\begin{equation}
F_W (z) = \frac{1}{\left(2 \pi \right)^{2N}} \int da \hspace{0.2 cm} \tilde F_W (a) e^{- i a \cdot z}.
\end{equation}
Here $a= (u,v)$ lives in the dual of the phase space. The function $\tilde F ( \tilde a)$, with $\tilde a= (u, - v)$ is known as the symplectic Fourier transform or chord function and finds many applications in the context of deformation quantization and decoherence \cite{Almeida}-\cite{Narcowich2}. 

If we substitute (10) into (1), we obtain
\begin{equation}
A(t) = \left. \tilde F_W \left( \frac{1}{i} \frac{\partial}{\partial z} \right) A_W (z,t) \right|_{z=0}.
\end{equation}
Equation (11) has a nice interpretation in the context of deformation quantization. In eq. (1) the objects appearing on the right-hand side are defined up to an isomorphism, in the following sense. The Weyl symbol stems from a correspondence rule according to which operators are first written in a fully symmetric form - the Weyl order - before they are 
``dequantized". Alternatively, one may choose other ordering prescriptions for the operators (e.g. normal ordering). The price to pay is that one also needs to change the corresponding quasi-distribution, in order to leave the expectation value (1) unchanged. This ambiguity has been systematized by Cohen \cite{Cohen}. Each correspondence rule is associated with an analytic so-called Cohen function $f(\xi)$, such that $f(0)=1$, which allows us to define a new $f$-symbol \cite{Lee}
\begin{equation}
A_f (z,t) = f \left( \frac{1}{i} \frac{\partial}{\partial z} \right) A_W (z,t)\,,
\end{equation}
and the corresponding quasi-distribution
\begin{equation}
F_f (z) = f^{-1} \left( i \frac{\partial}{\partial z} \right) F_W (z)\,.
\end{equation}
Likewise one naturally defines a $\star_f$-product and an $f$-bracket
\begin{equation}
A_f \star_f B_f = f \left( \frac{1}{i} \frac{\partial}{\partial z} \right) A_W \star_W B_W \,, \hspace{0.5 cm} \left[A_f, B_f \right]_f = f \left( \frac{1}{i} \frac{\partial}{\partial z} \right) \left[A_W, B_W \right]_W \,.
\end{equation} 
The dynamics is then dictated by
\begin{equation}
\dot A_f (z,t) = \left[H_f (z), A_f (z,t) \right]_f \,.
\end{equation}
Equation (11) can thus be interpreted in the following terms. If we regard $\tilde F_W$ as one of Cohen's functions, then $A_{\tilde F} (z,t) \equiv \tilde F_W \left( \frac{1}{i} \frac{\partial}{\partial z} \right) A_W (z,t)$ is just the $\tilde F_W$-symbol associated with the operator $\hat A$. The only difference is that one eventually sets $z=0$. The symbol $A_{\tilde F} (t)$ is a solution of eq.(15).

Note that eq.(11) remains valid even if $F_W$ is the Wigner function of a mixed state given by a density matrix $\rho$. However, in this paper we will be concerned with the pure states only.

\section{Coherent states}
   
The equation (11) derived in the previous section simplifies drastically if we choose the state $\psi$ to be the coherent state
\begin{equation}
\psi_{\alpha_0} (q) = \left( \frac{m \omega}{\pi \hbar} \right)^{\frac{N}{4}} \exp \left[ - \frac{m \omega}{2 \hbar} (q-q_0)^2 + \frac{i p_0}{\hbar} \cdot \left(q- \frac{q_0}{2} \right) \right]\,,\label{cswf}
\end{equation}
where in the standard notation
\begin{equation}
\alpha_0 \equiv \sqrt{\frac{m \omega}{2 \hbar}} q_0 + \frac{i p_0}{\sqrt{2 m \omega \hbar}}\,.
\end{equation}
>From (9) we have
\begin{equation}
\tilde F_W (u,v) = \exp \left[- \frac{m \omega \hbar u^2}{4} - \frac{\hbar v^2}{4 m \omega} + i u \cdot p_0 + i v \cdot q_0 \right].
\end{equation}
Substituting into (11), we obtain
\begin{equation}
\begin{array}{c}
A_{\alpha_0} (t) = \left. \exp \left[ \frac{\hbar}{4 m \omega} \frac{\partial^2}{\partial q^2} + \frac{\hbar m \omega}{4} \frac{\partial^2}{\partial p^2} \right] A_W (p+p_0, q+ q_0 , t) \right|_{q=p=0} = \\
\\
= \exp \left[ \frac{\hbar}{4 m \omega} \frac{\partial^2}{\partial q_0^2} + \frac{\hbar m \omega}{4} \frac{\partial^2}{\partial p_0^2} \right] A_W (p_0,q_0 ,t) \,.
\end{array}
\end{equation}
The last expression is well known to be the phase space symbol stemming from normal ordering \cite{Lee}. Now, this equation is particularly well suited for semiclassical expansions in powers of $\hbar$. Following the remark after the equation (\ref{six}), we conclude that if the Hamiltonian is of the form (\ref{seven}), then we get at the order $\hbar$ (cf.(8,19))
\begin{equation}
A_{\alpha_0} (t) = A_{cl} (p_0, q_0, t) + \left[ \frac{\hbar}{4 m \omega} \frac{\partial^2}{\partial q_0^2} + \frac{\hbar m \omega}{4} \frac{\partial^2}{\partial p_0^2} \right] A_{cl} (p_0, q_0,t) +  O (\hbar^2)\,.
\label{csf}\end{equation}
The expression (\ref{csf}) is valid for any trace-class operator $\hat A$ and any Hamiltonian of the form (\ref{seven}) provided the wavefunction is the coherent state (\ref{cswf}).

In fact these semiclassical expansions can be performed for other states, by using (11), as long as the chord function admits a regular expansion in powers of $\hbar$
\begin{equation}
\tilde F_W (a) = \sum_{n=0}^{\infty} \rho_n (a) \hbar^n \quad.
\end{equation}

In most applications one is interested in the case $\hat A = \hat q$. For simplicity, we shall henceforth consider a one-dimensional system. Let us define $Q(p_0, q_0,t) \equiv < \psi_{\alpha_0} | \hat q | \psi_{\alpha_0} >$ and $q_{cl} (p_0, q_0,t) \equiv q (p_0, q_0,t)$. In order to compare to the effective action approach, let us derive the equations of motion to order $\hbar$ for $Q(p_0, q_0,t)$. From (20) we have
\begin{equation}
\begin{array}{c}
m \ddot Q = m \ddot q + \frac{\hbar}{4 \omega} \frac{\partial^2}{\partial q_0^2} \ddot q + \frac{\hbar m^2 \omega}{4} \frac{\partial^2}{\partial p_0^2} \ddot q +  O (\hbar^2) = \\
\\
= - U'(q) - \frac{\hbar}{4 m \omega} \frac{\partial^2}{\partial q_0^2} U'(q) - \frac{\hbar m \omega}{4} \frac{\partial^2}{\partial p_0^2} U'(q) + O (\hbar^2)\,.
\end{array}
\end{equation} 
Notice that $U(q)$ depends on $p_0,q_0$ only through $q (p_0,q_0,t)$. Consequently
\begin{equation}
\begin{array}{c}
m \ddot Q = - U'(q) - \frac{\hbar}{4 m \omega} \left[ U'''(q) \left(\frac{\partial q}{\partial q_0} \right)^2 + U'' (q) \frac{\partial^2 q}{\partial q_0^2} \right] \\
\\
- \frac{\hbar m \omega}{4} \left[ U'''(q) \left(\frac{\partial q}{\partial p_0} \right)^2 + U'' (q) \frac{\partial^2 q}{\partial p_0^2} \right] +  O (\hbar^2)\,.
\end{array}
\end{equation}

We then obtain
\begin{equation}
U'(q) = U'(Q)  - U'' (q) \left[ \frac{\hbar}{4 m \omega} \frac{\partial^2 q}{\partial q_0^2} + \frac{\hbar m \omega}{4} \frac{\partial^2 q}{\partial p_0^2} \right]  +  O (\hbar^2)\,.
\end{equation}
By substituting (24) into (23), we finally obtain
\begin{equation}
m \ddot Q = - U'(Q) - \frac{\hbar}{4}  U'''(Q) \left[ \frac{1}{m \omega} \left(\frac{\partial Q}{\partial q_0} \right)^2 +  m \omega  \left(\frac{\partial Q}{\partial p_0} \right)^2  \right] +  O (\hbar^2)\,.
\end{equation}

This is a partial differential equation for $Q(p_0,q_0,t)$ where $(p_0,q_0)$ play a double role. They are variables, but also the initial conditions
\begin{equation}
\left. Q(p_0,q_0,t) \right|_{t=0} = q_0, \hspace{0.5 cm} \left. \dot Q(p_0,q_0,t) \right|_{t=0} = \frac{p_0}{m}\,.
\end{equation}

One may instead derive an ordinary differential equation for $Q$ and its time derivatives which is more suited for comparison with the effective action formalism. In order to do this let us consider
the solution for $q$ of the classical equations of motion given by
\begin{equation}
t = \pm \int_{q_0}^q dx \, \left[\frac{2}{m} \left( E - U(x) \right) \right]^{-1/2}\,,
\end{equation}
where
\begin{equation}
E= \frac{p_0^2}{2m} + U(q_0) = \frac{1}{2} m \dot q^2 + U(q)
\end{equation}
is the total energy of the classical system. Taking into account that $E$ and $q$ depend on $p_0$ and $q_0$, and applying the implicit function theorem we obtain
\begin{equation}
\frac{\partial q}{\partial q_0} = \frac{m}{p_0} \dot q + \frac{\dot q U'(q_0)}{2} \sqrt{\frac{m}{2}} f(q,p_0,q_0)\quad,\quad
\frac{\partial q}{\partial p_0} = \frac{p_0 \dot q }{2\sqrt{2m}} f(q,p_0,q_0)\,,
\end{equation}
where
\begin{equation}
f(q,p_0,q_0) \equiv \int_{q_0}^q dx \hspace{0.2 cm} \left[E- U(x) \right]^{- 3/2}\,.
\end{equation}

In the eq.(25) it is immaterial whether we write $q$ or $Q$ in the terms proportional to $\hbar$. From (29) we then obtain
\begin{equation}
\begin{array}{c}
m \ddot Q = - U' (Q) - \frac{\hbar}{4} \dot Q^2 U'''(Q) \left\{\frac{m}{\omega p_0^2} + \frac{V'(q_0)}{\omega p_0} \sqrt{\frac{m}{2}} f(Q, p_0 , q_0) + \right.\\
\\
\left. + \left[\frac{\left(U' (q_0) \right)^2}{8 \omega} + \frac{\omega p_0^2}{8} \right] f^2 (Q, p_0 , q_0) \right\} + O (\hbar^2)\,.
\end{array}
\end{equation}

If we neglect the $O(\hbar^2)$ terms in (31) we obtain a second order nonlinear ordinary differential equation for $Q(t)$ with the initial conditions (26).

\section{The anharmonic oscillator}

As an application we consider the quartic anharmonic oscillator
\begin{equation}
H= \frac{p^2}{2m} + \frac{1}{2} m \omega^2 q^2 + \frac{\l}{4!}\, q^4 \quad \,,
\end{equation}
where $\l$ is a positive coupling constant. Let us define $\L =\l/4!$ and
\begin{equation}
\alpha^2 \equiv \frac{\sqrt{m^2 \omega^4 + 16 E \L} - m \omega^2}{4 \L}\quad,\quad
\beta^2 \equiv \frac{\sqrt{m^2 \omega^4 + 16 E \L} + m \omega^2}{4 \L} \quad.
\end{equation}
By substituting 
\begin{equation}
x= \frac{\alpha \beta \sin \varphi}{\sqrt{\beta^2 + \alpha^2 \cos^2 \varphi}}\quad,
\end{equation}
into (30) we obtain
\begin{equation}
\begin{array}{c}
f(q,p_0,q_0) = \frac{1}{\left[\L ( \alpha^2 + \beta^2 ) \right]^{3/2}} \left\{ \frac{1}{\beta^2} E \left( \left. \varphi (q) \right| n \right) + \right. \\
\\
\left. + \frac{1}{\alpha^2} \left[ \tan \varphi (q) \sqrt{1- n \sin^2 \varphi (q)} +  F \left( \left. \varphi (q) \right| n \right) - E \left( \left. \varphi (q) \right| n \right) - \left( q \longleftrightarrow q_0 \right) \right] \right\},
\end{array}
\end{equation}
where
\begin{equation}
\varphi (q) \equiv \arcsin \sqrt{\frac{q^2}{n (q^2 + \alpha^2)}}, \hspace{1 cm} n \equiv \frac{\alpha^2}{\alpha^2 + \beta^2}, \hspace{0.5 cm} 0 \le n \le 1,
\end{equation}
and 
\begin{equation}
F \left( \phi | m \right) \equiv \int_0^{\phi} d \theta \, \frac{1}{\sqrt{1 - m \sin^2 \theta}}\quad,
\quad
E \left( \phi | m \right) \equiv \int_0^{\phi} d \theta \, \sqrt{1 - m \sin^2 \theta} \,,
\end{equation}
are the incomplete elliptic integrals of the first and second kind, respectively \cite{Abramowitz}.

In order to compare this with the equations of motion stemming from the effective action, let us expand the right-hand side of (31) in powers of $\l$
\bea
\ddot Q &=& - \,\omega^2 Q - \frac{ \l}{6m} Q^3 \nonumber\\
&\,&- \, \hbar \l \frac{k_t^2 x_t}{4m^2} \left\{\frac{1}{\omega p_0^2} + \frac{m \omega  q_0}{p_0 \sqrt{2m}} f_0 (x_t, p_0,q_0) + \frac{\omega E_0}{4} f_0^2 (x_t, p_0 ,q_0 ) \right\}\nonumber\\
&\,& +\,O(\hbar\l^2)\,, \label{acsem}
\eea
where $x_t$ and $k_t$ are the classical harmonic oscillator (HO) solutions for the coordinate and the momentum
\begin{equation}
x_t = q_0 \cos (\omega t) + \frac{p_0}{m \omega} \sin (\omega t)\quad,\quad k_t = p_0 \cos (\omega t) - m \omega q_0 \sin (\omega t)\,,
\end{equation}
with the energy
\begin{equation}
E_0 = \frac{p_0^2}{2m} + \frac{1}{2} m \omega^2 q_0^2 \quad.
\end{equation}

Moreover
\begin{equation}
f_0 (x_t, p_0 ,q_0) = \int_{q_0}^{x_t} dx \hspace{0.2 cm} \left( E_0 - \frac{1}{2} m \omega^2 x^2 \right)^{- 3/2}= \frac{\sqrt{2m}}{E_0} \left( \frac{x_t}{k_t} - \frac{q_0}{p_0} \right).
\end{equation}
The terms proportional to $\hbar \l$ in (38) yield
$- \frac{\hbar \l x_t}{4m^2 \omega}$ and
up to order $\hbar \lambda$ we may replace $x_t$ by $Q$ in the terms proportional to $\hbar \lambda$. By neglecting the $O(\hbar\l^2)$ terms we obtain
\begin{equation}
\ddot Q + \left(\omega^2 + \frac{\hbar\l }{4m^2 \omega}\right) Q + \frac{\l }{6m} Q^3 =0\quad,\label{csfo}
\end{equation}
as a semiclassical equation of motion.

\section{The effective action results}

Let us consider a $D$-dimensional scalar field theory given by the action
\be S[\f] = \int d^D x \,  \left[ \ha(\pr_\mu \f)^2 - \ha M^2 \f^2 - V(\f) \right]\quad,\ee
where $V(\f)$ is polynomial in $\f$.
The corresponding effective action can be written as \cite{ram}
\be \G [\f] = \sum_{n=2}^\infty \int d^D x_1 \cdots \int d^D x_1 \frac{1}{n!}\G^{(n)} (x_1,\cdots,x_n)\f(x_1)\cdots\f(x_n) \quad,\ee
where
\be \G^{(n)} (x_1,\cdots,x_n) = \int d^D p_1 \cdots\int d^D p_n e^{i(p_1 x_1 + \cdots +p_n x_n )}\left(\G^{(n)}(p_1, \cdots, p_n)\right)^{\e_n}\quad.\ee
The $\G^{(n)}(p)$ are the momentum space $n$-particle irreducible Greens functions whose external legs are amputated for $n>2$ and $\e_n = -1$ for $n=2$ and $\e_n =1$ for $n>2$. These objects can be calculated perturbatively via the Feynman diagrams and for $V=\l\f^4 /4!$ the perturbative expansion can be organized in powers of $\l$ (number of the vertices) and in powers of $\hbar$ (number of the loops), so that
\be \G^{(n)}(p) = \sum_{v,l} \l^v \,\hbar^l \,\G^{(n)}_{v,l}(p) \quad.\ee

Up to the order of $\l \hbar$ only the tadpole diagram contributes \cite{ram}, so that in the case of $D=1$ field theory, i.e. Quantum Mechanics (QM), one obtains 
\bea \G_E^{(2)}(p)&=&{1\over p^2 + M^2} - {\l\hbar/2\over (p^2 + M^2 )^2}\int_{-\infty}^{\infty} {dq\over 2\pi} {1\over q^2 + M^2} + O(\l^2\hbar)\,\nonumber\\
&=& {1\over p^2 + M^2} - {\l\hbar\over 2 (p^2 + M^2 )^2}{1\over 2M}+ O(\l^2\hbar)\quad,\eea
where $\G_E$ is the Euclidean propagator. The physical (Minkowski) propagator is given by $\G^{(2)}(p)=-\G_E^{(2)}(ip)$, and the position space vertex function is given by
\be \G^{(2)}(x_1,x_2)= 2\pi \int {dp_1 \over 2\pi} \int {dp_2 \over 2\pi}\, \d(p_1+p_2)\left(\G^{(2)}(p_1)\right)^{-1} e^{i(p_1 x_1 + p_2 x_2)}\quad.\ee
This gives
\be \G [\f]= S[\f] + {\l\hbar\over 8M}\int_{x_1}^{x_2} dx\, \f^2(x) + O(\l^2 \hbar) \quad. \ee 
Passing to the anharmonic oscillator parameters ($M\to\o\,,\,\l\to\l/m^2$) gives
\be \G [q]= S[q] + {\l\hbar\over 8m^2\o} \int_{t_1}^{t_2}  dt\,q^2(t)+ O(\l^2 \hbar) \quad. \ee
Up to order $O(\hbar\l)$ this action gives the same equation of motion as (\ref{csfo}).

When going to higher orders in perturbation theory, one obtains
the non-local terms in the effective action. For example, at the order $\l^2\hbar$ one has to include four $\G^{(4)}$ diagrams (one three-diagram plus three one-loop diagrams \cite{ram}) so that
\bea \G_E^{(4)}(p)&=&-\l + \l^2 \hbar  \int_{-\infty}^{\infty} {dq\over 2\pi} {1\over q^2 + M^2}{1\over (p_1+p_2 -q)^2 + M^2 } \nonumber\\
&+& \l^2 \hbar  \int_{-\infty}^{\infty} {dq\over 2\pi} {1\over q^2 + M^2}{1\over (p_1-p_3 +q)^2 + M^2 }\nonumber\\&+&\l^2 \hbar  \int_{-\infty}^{\infty} {dq\over 2\pi} {1\over q^2 + M^2}{1\over (p_1-p_4 +q)^2 + M^2 }\,.\eea
This gives
\bea \G_E^{(4)} (p) = -\l &+& {\l^2 \hbar\over M}{\Big[}\, -{1\over (p_1 +p_2)^2 + 4M^2 }\nonumber\\ 
&+&{1\over (p_1-p_3 )^2 + 4M^2 }+{1\over (p_1-p_4)^2 + 4M^2 }\,{\Big]}\quad.\eea

The physical vertex function is given by $\G^{(4)}(p)=\G_E^{(4)}(ip)$, and the position space vertex function is given by
\bea \G^{(4)}(x_1,\cdots,x_4)&=& 2\pi \int {dp_1 \over 2\pi}\cdots \int {dp_4 \over 2\pi}\, \d(p_1+\cdots+p_4)\nonumber\\ &\,&\G^{(4)}(p_1,\cdots,p_4)\, e^{i(p_1 x_1 +\cdots+ p_4 x_4)}\quad.\eea
One then obtains the effective action contribution 
\be \int dx \left[ -{\l\over 4!}\f^4 - {\l^2\hbar\over 4! M}\int dy \,G_{2M}(x-y)\,\f^2 (x)\f^2 (y)\right]\quad,\ee
where
\be G_\mu (x) =\,Re\,\int {dp\over 2\pi}{e^{ipx}\over p^2 -\mu^2 +i\e} ={1\over 2\mu}\sin(\mu x)[\theta(-x) -\theta(x)] \quad,\ee
is the real part of the $D=1$ Feynman propagator and $\th(x)=1$ for $x> 0$ and $\th(x)=0$ for $x\le 0$. 

By going to the QM parameters via $\l\to\l/m^2$, $M\to\o$ and $\f\to\sqrt{m}q$, one obtains the following 
effective equations of motion
\bea 0 &=& \ddot{q} + \o^2 q + \frac{\l}{6m}\, q^3 + {\l\hbar\over 4m^2\o}\,q \nonumber\\
&\,&+ \,{\l^2\hbar\over 6 m^3\o}\,q\,\int_{-\infty}^{\infty} d\t \, G_{2\o} (t-\t)q^2 (\t)\,+\, O(\l^3 \hbar)\,.\label{peom}\eea
This effective equation of motion can be solved perturbatively as
\be q(t) = q_{0,0} (t) + \sum_{m\ge 1,\,n\ge 0}\l^m \hbar^n q_{n,m} (t) \quad,\label{peas}\ee
where $q_{0,0} (t)$ is the classical HO solution. The first and the second quantum correction will satisfy the equations
\bea \ddot{q}_{1,1} + \o^2 q_{1,1} &=& -{1\over 4m^2\o}\,q_{0,0} \label{eafc}\\
\ddot{q}_{1,2} + \o^2 q_{1,2} &=& -\frac{1}{2m}\, q_{0,0}^2 q_{1,1} -{1\over 4m^2\o}\,q_{0,1}\nonumber\\ &\,&-\,{1\over 6 m^3\o}\,q_{0,0}\,\int_{-\infty}^{\infty} d\t \, G_{2\o} (t-\t)q_{0,0}^2 (\t) \quad.\eea

The solution (\ref{peas}) has a classical part given by
\be q_c = q_{0,0} + \l q_{0,1}+ \l^2 q_{0,2}+ \cdots \quad.\label{clex}\ee
Up to the order of $\l^2$ the classical solution is given by
\bea q_{0,0} &=& a \cos\o t + b\sin\o t \label{hos}\\
q_{0,1} &=& a_1 \cos 3\o t + b_1 \sin 3\o t + (c_1 t +e_1)\cos\o t + (d_1 t +f_1)\sin\o t \label{fcc} \\
q_{0,2} &=& a_2 \cos 5\o t + b_2 \sin 5\o t + (c_2 t +c'_2)\cos 3\o t + (d_2 t + d'_2)\sin 3\o t \nonumber\\
&\,& + (e_2 t^2 + e'_2 t +g_2)\cos\o t + (f_2 t^2 + f'_2 t +h_2)\sin\o t \,,\label{scc}\eea
where $a=q_0$, $b=p_0/m\o$ and the coefficients $a_k$, $b_k$, ... are the homogeneous polynomials of $a$ and $b$ of the order $2k+1$, see the equations (A.4) and (A.5) in the Appendix. 

From (\ref{hos}) and (\ref{eafc}) it follows that
\be q_{1,1} =  {t\over 8m^2\o^2}(b\cos\o t - a \sin\o t )-{b\over 8m^2 \o^3}\sin\o t \quad,\label{csi}\ee
which coincides with the coherent state result
\be Q_{1,1} = \frac{1}{4m\o}\left(\frac{\pr^2}{\pr a^2} + \frac{\pr^2}{\pr b^2}\right)q_{0,1} \quad.\ee
This is an expected result since the equations of motion coincide up to this order. 

In order to solve the equation for $q_{1,2}$ we need to evaluate the expression
\bea I(t)&=& \int_{-\infty}^{\infty}d\t\,G_{2\o}(t-\t) [a\cos(\o \t)+b\sin(\o\t)]^2 \nonumber\\
 &=&  {\xi\over 32\o^2}\left[(b^2 - a^2)\cos(2\o t) -2ab\sin(2\o t)-4(a^2 + b^2)\right]\quad,\eea
where $\xi=\int_0^\infty dx\,\sin x $. The divergent integral can be regularized by the formula 
\be \int_0^\infty dx \,e^{-\e x + i x} = (\e -i)^{-1} \quad,\quad\e>0\quad,\ee
which in the $\e\to 0$ limit gives $\xi=1$. The solution will then have the form 
\bea q_{1,2} &=& (\g_2 t +\g'_2)\cos 3\o t + (\d_2 t + \d'_2)\sin 3\o t \nonumber\\ 
&\,& +\, (\e_2 t^2 + \e'_2 t + \k_2)\cos\o t + (\f_2 t^2 + \f'_2 t +\c_2)\sin\o t \,,\label{eaii}\eea
see the equation (A.7).

The coherent state expectation value will give the correction of the form
\bea  Q_{1,2} &=& \frac{1}{4m\o}\left(\frac{\pr^2}{\pr a^2} + \frac{\pr^2}{\pr b^2}\right)q_{0,2} \nonumber\\
              &=& ( \tilde c_2 t + \tilde c'_2)\cos 3\o t + 
              ( \tilde d_2 t +  \tilde d'_2)\sin 3\o t \nonumber\\
&\,& +\, (\tilde e_2 t^2 + \tilde e'_2 t +\tilde g_2 )\cos\o t + (\tilde f_2 t^2 + \tilde f'_2 t +\tilde h_2 )\sin\o t \,,\label{csii}\eea
see the eq. (A.6), which could in principle coincide with (\ref{eaii}) if the corresponding coefficients were identical. However, by comparing (A.6) to (A.7) one can see that $q_{1,2}\ne Q_{1,2}$, and therefore there is a discrepancy at the order $\l^2 \hbar$. 

One can also try to obtain a nonperturbative in $\l$ effective action equations of motion, which amounts to suming all the diagrams with different powers of $\l$ at a fixed order in $\hbar$.
This can be achieved by using the saddle point approximation in the path-integral formalism, see \cite{ram}. This then boils down to evaluating the traces of $D$-dimensional differential operators. The drawback of this approach is that one can only obtain certain terms at a fixed order of $\hbar$, i.e. not the complete correction. The standard approximation is
\be \G [\f] \approx \int d^D x \left[ \ha (1+Z(\f))(\pr_\mu \f)^2 -\ha M^2 \f^2 -V_{eff}(\f)\right]\quad,\label{epa}\ee
where $V_{eff}$ is the effective potential. In the $D=1$ case one obtains at $O(\hbar)$ \cite{jl3} 
\be V_{eff}(q) = V(q)+{\hbar\o\over 2}\left(\sqrt{1+ {V^{\prime\prime}(q)\over m\o^2}} - 1\right)\,,\,
Z(q) = {\hbar\over 32 m^3}{\left(V^{\prime\prime\prime}(q)\right)^2\over \left(\o^2 + 
{ V^{\prime\prime}(q)\over m}\right)^{5/2}}\,,\ee
so that
\be \G [q] \approx \int_{t_1}^{t_2} dt \left[ \ha m(1+Z(q))(\dot q)^2 -\ha m\o^2 q^2 -V_{eff}(q)\right]\quad.\label{efpa}\ee

The equations of motion coming from (\ref{efpa}) are given by
\be 0=(1+Z)\ddot q + \ha Z' (q) (\dot q)^2 + \o^2 \, q  + V'_{eff}(q)\quad.\label{efpe}\ee
By using
\bea V_{eff} (q) &=& \l {q^4 \over 4!} + \hbar\l { q^2 \over 8m\o}- \hbar \l^2{ q^4 \over 64 m^2 \o^3} + O(\l^3 \hbar) \quad,\\
 Z(q) &=& \hbar\l^2 {q^2 \over 32 m^3 \o ^5} + O(\l^3 \hbar) \quad,\eea
(\ref{efpe}) can be expanded in powers of $\l$ and one
can see that (\ref{efpe}) agrees with (\ref{peom}) up to $O(\l\hbar)$. At $O(\l^2\hbar)$ there is a disagreement, reflecting the fact that there are other terms contributing at $O(\hbar)$ which we have not included in the approximation (\ref{efpa}) (for example $\int dt A(q)(\dot q)^4$ or non-local terms).

One can also compare the perturbative solutions, and it can be shown by using the techniques from the Appendix that the $O(\hbar\l^2)$ perturbative solution of (\ref{efpe}) has a form
\bea \tilde q_{1,2} &=& \tilde\g'_2 \cos 3\o t + \tilde\d'_2 \sin 3\o t \nonumber\\ 
&\,& +\, (\tilde\e_2 t^2 + \tilde\e'_2 t +\tilde\k_2)\cos\o t + (\tilde\f_2 t^2 + \tilde\f'_2 t +\tilde\c_2)\sin\o t \,,\label{efps}\eea
which differs from the coherent state and the perturbative effective action results by the absence of the $t\cos 3\o t$ and $t\sin 3\o t$ terms.

\section{Conclusions}

Comparing the effective action equations of motion (\ref{peom}) and (\ref{efpe}) to the coherent state expectation value equation of motion (\ref{acsem}) gives an agreement up to $O(\l\hbar)$. For the higher orders it is simpler to compare the solutions and we find that at $O(\l^2 \hbar)$
there are discrepancies between all three approaches. Although all three solutions (\ref{eaii}), (\ref{csii}) and (\ref{efps}) are linear combinations of $t^n \cos (2k+1)\o t$ and  $t^n \sin (2k+1)\o t$ functions, the corresponding coefficients differ. 

Note that the effective action perturbative solution (\ref{eaii}) depends on a regularization dependent parameter $\xi$. However, the $\xi$-independent coefficients are different from the corresponding expectation value solution coefficients, see the Appendix. Hence the discrepancy cannot be explained as a regularization scheme artifact. Therefore the state whose EV gives the effective action trajectory is not the coherent state in the AHO case. This agrees with the assumption made in the reference \cite{br}, where the effective action state was taken to be the ground state of the AHO with a source. The arguments for this state were first presented in \cite{jl2}. 

The discrepancy between the perturbative effective action solution (\ref{eaii}) and the perturbative effective potential solution
(\ref{efps}) is simply due to the fact that the effective potential method is an approximation which did not take into account all possible terms which could contribute at a given order.

Although the formula (\ref{csf}) is valid for the coherent states, it can be used as an approximation for the field expectation value coming from the effective action. We expect that it can give the same result as the effective action in the case of two-dimensional integrable field theories. This happens in the case of the Callan-Giddings-Harvey-Strominger dilaton gravity model \cite{m}, and this is not surprising because the
two-dimensional integrable field theories are closely related to the free-field theories for which the effective action and the coherent state expectation value give the same dynamics.  

In order to use the formula (\ref{csf}) in the field theory case, one would have to extend the DQ formalism to the field theory case, see \cite{ftdq}.  

Our results imply that the coherent state expectation values define in general an alternative semiclassical dynamics to that coming from the effective action. The advantage of using the coherent states is that they give directly a semiclassical trajectory through computable $\hbar^n$ corrections to the classical one, while in the effective action approach it is often impossible to obtain a complete $\hbar^n$ correction to the classical equations of motion. Furthermore, the effective action equations of motion are non-local and difficult to solve nonperturbatively in the coupling constant. Hence using the coherent states to obtain the semiclassical trajectories represents a promissing approach.

\bigskip
\noindent{\bf APPENDIX}

\bigskip
The classical AHO equation of motion can be solved perturbatively via the expansion (\ref{clex}). One then obtains the equations
$$ L q_{0,1} = -{q_{0,0}^3 \over 6m} \quad,\quad L q_{0,2} = -{q_{0,0}^2 q_{0,1} \over 2m}\quad,\quad\cdots\eqno(A.1)$$
where $L=\frac{d^2}{dt^2} + \o^2$. These equations can be solved by the method of undetermined coefficients
for the forced HO equation of motion $Lq = f(t)$. 

In order to solve the first two equations in (A.1) we will need a particular solution for
$$f(t) = (At +A')\cos\O t + (Bt + B')\sin\O t \quad.$$
It is given by
$$  q(t) = \left({At + A'\over \o^2 -\O^2} -{2B\O\over (\o^2 -\O^2 )^2}\right)\cos\O t + 
\left({Bt + B'\over \o^2 -\O^2} +{2A\O\over (\o^2 -\O^2 )^2}\right)\sin\O t \,, \eqno(A.2)$$
for $\O\ne\o$ and by
$$ q(t) = {1\over 4\o}\left(-Bt -2B' +{A\over\o}\right)t\cos\o t + 
{1\over 4\o}\left(At +2A' +{B\over\o}\right)t\sin\o t \quad, \eqno(A.3)$$
for $\O=\o$.

By using the formulas (A.2) and (A.3) we obtain (\ref{fcc}) and (\ref{scc}). The initial conditions
$$ q(0) = a \quad,\quad \dot q (0) = \o b \quad,$$
are imposed by requiring
$$ q_{n,m}(0) = 0 \quad,\quad \dot q_{n,m} (0) = 0 \quad,$$
for $(n,m)\ne (0,0)$. These conditions determine the coefficients of the HO terms as
$$ e_1 = -a_1 \,,\, f_1 = -{c_1 \over\o} \,,\, g_2 = -a_2 -c'_2 \,,\, h_2 = -5b_2 -{c_2 + e'_2 \over \o} \,,$$
etc.

One then obtains
$$ q_{0,1} = {1\over 192 m\o^2}{\Big \{}a(a^2 - 3b^2)\cos 3\o t - b(b^2 -3a^2 )\sin 3\o t $$
$$ + [-a^3 + 3ab^2 +12b(a^2 + b^2 )\o t ]\cos\o t $$
$$+ [21 a^2 b + 9b^3 + 4a(a^2 + b^2 )\o t]\sin\o t {\Big \}}\quad,\eqno(A.4)$$
and
$$ q_{0,2} = {1\over 36864 m^2 \o^4}{\Big \{}(a^5 -10a^3 b^2 +5ab^4 )\cos 5\o t$$
$$ -12[2a^5 -15a^3 b^2 -(9a^4 b +6a^2 b^3 -3b^5 )\o t]\cos 3\o t $$
$$+[23a^5 -170a^3 b^2 -113ab^4 -48b(7a^4 + 19a^2 b^2 +8b^4 )\o t - 72a(a^2 + b^2 )^2 \o^2 t^2 ]\cos\o t$$
$$+(b^5 -10 a^2 b^3 + 5a^4 b)\sin 5\o t$$
$$ +[132a^4 b +12a^2 b^3 +48b^5 -(36a^5 -72a^3 b^2 -108ab^4 )\o t ]\sin 3\o t $$
$$+[599a^4 b +854a^2 b^3 +271b^5 +(96a^5 + 528a^3 b^2 +240ab^4 )\o t$$
$$ +(72a^4 b -144a^2 b^3 -72b^5 )\o^2 t^2 ]\sin \o t {\Big \}}\,.\eqno(A.5)$$

The coherent state expectation value is then obtained from (\ref{csf}). This gives (\ref{csi}) and
$$ Q_{1,2} = {1\over 6144 m^3 \o^5}{\Big\{} [24(3 a^2 b - b^3)\o t + 99ab^2 -5a^3 ]\cos 3\o t $$
$$-[63a^2 b -41 b^3 + 24(a^3 -3ab^2)\o t ]\sin 3\o t $$
$$ +[5a^3 - 99ab^2 -396b (a^2 + b^2)\o t -72a(a^2 +b^2)\o^2 t^2 ]\cos\o t $$
$$+ [27b (19a^2 + 11b^2)+ 4a(31a^2 + 63 b^2) \o t -72b(a^2 +b^2 )\o^2 t^2 ]\sin\o t \Big{\}}\,.\eqno(A.6) $$

The $O(\hbar\l^2)$ correction from the effective action equation of motion (\ref{peom}) can be obtained by solving the equation
\bea Lq_{1,2} &=& -{1\over 2m} q_{0,0}^2 q_{1,1} - {1\over 4m\o^2} q_{0,1} \nonumber\\
&\,&+\, {\xi\over 96m^3 \o^2}\left( 2a^2 + 2b^2 + {a^2 - b^2 \over 2}\cos 2\o t + ab \sin\o t \right)q_{0,0} \quad,\nonumber\eea
where $\xi$ is a regularization dependent constant. The Cauchy problem solution will have the form (\ref{eaii}), and the
$\xi$ independent coefficients are $\g_2 , \d_2 , \e_2$ and $\f_2$. For $\xi =1$ one obtains
$$ q_{1,2}={1\over 3072m^3 \o^5}{\Big\{}[27ab^2 -5a^3 + 6(3a^2 b -b^3 )\o t ]\cos 3\o t  $$
$$ + [11b^3 -21a^2 b +6(3ab^2 - a^3)]\sin 3\o t$$
$$ +[5a^3 -27ab^2 -6b (21a^2 +25b^2 )\o t -24a(a^2 +b^2 )\o^2 t^2 ]\cos\o t$$
$$ +[3b(57a^2 +41b^2 ) +2a (31a^2 +51b^2) \o t -24b(a^2 +b^2 )\o^2 t^2 ]\sin\o t {\Big \}}\,.\eqno(A.7)  $$

We have checked the results (A.4-7) by using the MATHEMATICA program \cite{w}.

\bigskip
\noindent{\large{{\bf Acknowledgments}}} 

We would like to thank C. Presilla and M.J. Prata for useful suggestions.
This work has been supported by the FCT grant POCTI/MAT/45306/2002.

\end{document}